\documentstyle[twocolumn,aps,epsf]{revtex}

\def\myfigure#1#2{{\leftskip=0.000753\textwidth \rightskip\leftskip\small
\begin{figure}\baselineskip=14pt plus 2pt minus 1pt
\centerline{#1}\nobreak\smallskip\nobreak #2\end{figure}}}
\global\firstfigfalse

\draft	

\begin{document}
\title{Droplet Fluctuations in the Morphology and Kinetics of Martensites}

\author{Madan Rao$^{1}$\cite{MAD} and Surajit Sengupta$^2$\cite{SUR}}

\address{$^1$Institute of Mathematical Sciences, Taramani, Madras 600113, India\\
$^2$Material Science Division, Indira Gandhi Centre for Atomic 
Research, Kalpakkam 603102, India}

\date{\today}

\maketitle

\begin{abstract}
We derive a coarse grained, free-energy functional which describes 
droplet configurations arising on nucleation of a product crystal
within a parent. This
involves a new `slow' vacancy mode that lives at the parent-product interface.
A mode-coupling theory suggests that a {\it slow} quench from the parent
phase produces an equilibrium product, while
a {\it fast} quench produces a metastable martensite.
In two dimensions, the martensite nuclei grow as `lens-shaped'
strips having alternating twin domains, with well-defined front
velocities. Several empirically known structural and kinetic relations
 drop out naturally from our theory.

\end{abstract}


In modern metallurgical parlance, a {\it martensitic transformation}\cite{WWW},  has come to 
denote any diffusionless, structural transformation resulting in a 
long-lived
metastable phase with a high degree of short-range order.
 For instance,
when adiabatically cooled below $T_c = 910^{\circ}$C,
Fe undergoes
a first-order structural transition from an FCC solid ({\it austenite}) to an equilibrium BCC solid ({\it ferrite}). A faster quench produces instead, a rapidly transformed
metastable phase called the {\it martensite}. On nucleation,
martensite `plates' ($\sim\,1 \mu m$) grow with a constant front velocity ($\,\sim\!10^5 
cm\,s^{-1}$)\cite{NISHROIT}\,; fast compared to typical atomic diffusion speeds. The plates consist of an alternating array of twin
BCC crystals along the equatorial plane. Despite the considerable amount of empirical\cite{NISHROIT}
and theoretical work\cite{KRUM,SETH,GOOD}, a unified theoretical approach 
addressing both kinetics and morphology, and capable of describing the metastable martensite as well as the equilibrium ferrite, has not emerged.
In this Letter, we present a mode-coupling theory for the nucleation
and growth of a product crystalline droplet within a parent crystal.
We show that for slow quenches, the droplet grows diffusively as an equilibrium ferrite inclusion, while for fast quenches, the droplet grows ballistically, as a martensite having twinned internal substructure, with a speed comparable to the sound velocity.

Our mode-coupling dynamics involves ``slow'' modes
that change over time scales corresponding to the propagation of
the nucleation front. The slow modes of a solid undergoing a first-order
structural transformation are the displacement field ${\bf u}({\bf r},t)$, 
and the momentum density ${\bf g}({\bf r},t)$. Imagine, however, a droplet of the product nucleating within a parent crystal at time $t=0$. It is clear (Fig.\ 1(a)), that an atomic mismatch is generated
at the parent-product interface\cite{NISHROIT} as soon as the nucleus is formed. This mismatch appears as a discontinuity in the normal component of the
displacement field across the parent-product interface\cite{LL}
$\Delta {\bf u} \cdot {\hat {\bf n}} \equiv \phi$, and leads to a compression or dilation of the local atomic environment, Fig.\ 1(b).
Since the martensite front velocity is large
compared to atomic diffusion times, $\phi({\bf r},t)$ appears as a 
{\it new slow mode} (vacancy field), which relaxes 
diffusively over a time scale $\tau_{\phi}$.

\myfigure{\epsfysize1.6in\epsfbox{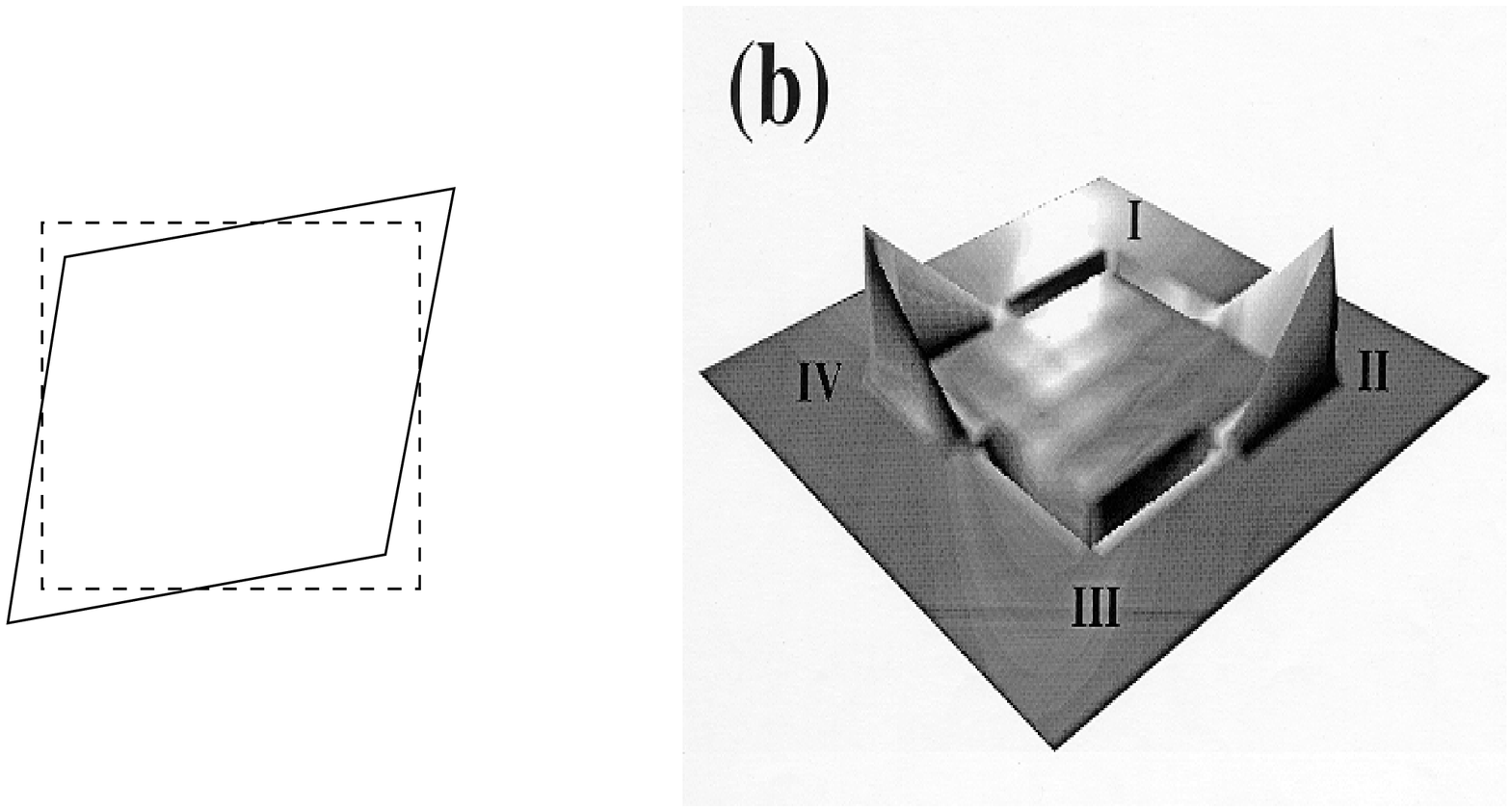}}{\vskip0inFIG.~1.~~
(a)~Inclusion of a product (rhombus) in a parent crystal, showing
the discontinuity in ${\bf u}$ across the parent-product interface at $t=0$.
(b)~A 3-d plot of the initial value of $\phi(x,y)$ for the inclusion shown in (a). The four corners in (a) and (b) are clearly marked.}

A Langevin description of the dynamics of the transformation, requires a free-energy functional {\it which describes all
intermediate configurations between an austenite and a ferrite}.
Thus the usual elastic free-energy functional of a solid, $F_{el}$, has to
be augmented by an interfacial term, 
describing the parent-product interface. As we see below, such
droplet configurations do not affect equilibrium behaviour, though
their effect on dynamics is significant.

Nucleation of a droplet of the product of size $L$ and interfacial thickness $\xi$, results in a strained crystal. The free-energy functional\cite{RY} of this strained crystal is given by,
\begin{equation}
{\cal {F}} = \sum_{\{{\bf R},{\bf R}^{\prime}\}} c\,(|{\bf R}-{\bf R}^{\prime}+{\bf u}({\bf R})-{\bf u}({\bf R}^{\prime})|)\,\,,
\label{eq:FREE}
\end{equation}
where $\{{\bf R}\}$ represent the lattice vectors of the parent crystal,
${\bf u}({\bf R})$ are the displacement fields and
$c(|{\bf r} - {\bf r}^{\prime}|)$ is the 
direct correlation function\cite{HM} of the liquid at freezing. 
The above expression is exact at $T=0$\,; corrections 
are of the order of the r.m.s. fluctuations of the atoms about their
perfect lattice positions, which are small in the solid phase\cite{MANG}.

Since the range of $c$ is of order $\xi$, 
the sum over $\{{\bf R},{\bf R}^{\prime}\}$ in Eq.\ (\ref{eq:FREE}), can be split into two parts, with $\{{\bf R},{\bf R}^{\prime}\}$ being on the {\it same
 or on either side} of the interface. 
Expanding ${\bf u}({\bf R}^{\prime})$ about ${\bf R}$ in the first
part, leads to the usual bulk elastic free-energy functional 
$F_{el}$\cite{MANG}. However, as argued above, the assumption of continuity
breaks down for the second part, and so 
{\it such an expansion cannot be carried out}.
In the limit $\xi/L \ll 1$, the interface can be parametrized by a sharp surface $\Gamma\,({\bf r})=0$, with ${\bf R}$ and ${\bf R}^{\prime}$ lying
infinitesimally close to $\Gamma =0$.
Across this interface, the normal component of ${\bf u}$ is 
discontinuous\,; $lim_{\left|{\bf R}-{\bf R}^{\prime}\right|
\to 0} ({\bf u}({\bf R}) - {\bf u}({\bf R}^{\prime}))\cdot
{\hat {\bf n}} \equiv ({\bf u}_{+} - {\bf u}_{-})\cdot{\hat {\bf n}} \equiv \phi$, where ${\hat {\bf n}}$ is the unit normal to the
interface. The discontinuity $\phi({\bf r})$, which is simply $\xi\,(\rho({\bf r})|_{\Gamma=0} - \bar{\rho})/\bar{\rho}$, is the local vacancy field 
($\rho$ and $\bar{\rho}$ are the local and the average densities respectively).  In the continuum limit, this leads to the free-energy functional (to leading order in the discontinuity),
\begin{equation}
{\cal {F}} = F_{el} + \frac{\gamma}{2\xi^2} \int \,\frac {d{\bf r}}{\Omega_0}\,\, [\,({\bf u}_{+} - {\bf u}_{-}\,)\cdot{\hat {\bf n}}\,]^2 \,\,\xi\,\delta(\Gamma)\,\,,
\label{eq:INT}
\end{equation} 
where $\Omega_0$ is the unit cell volume of the parent.
The coefficient $\gamma \equiv {\Omega_0}^{-1}\sum^{\prime}
 \xi^2 \partial_n \partial_n c(r)$, where the prime denotes a sum across the
interface, is clearly the
{\it surface compressibility modulus} of vacancies, 
whose magnitude is of the order of the
bulk elastic moduli\,; it is however dependent on the local orientation of the parent-product interface. Since the strains $\epsilon_{ij} \equiv (\partial_i u_j + \partial_j u_i)/2$
 are continuous across the interface, we replace the
delta function in Eq.\ (\ref{eq:INT}) by a regulator $\xi^{-1}\,[1-\exp(-\xi \, \partial_{n} \epsilon_{ij})^2]$, so that the final regulated free-energy functional
which incorporates {\it all the slow modes} in the problem is,
\begin{equation}
{\cal {F}} = F_{el} + \frac{\gamma}{2\Omega_0} \int \,d{\bf r}\,\, {\phi^{2}} \,\,({\partial_{n}}\,\, \epsilon_{ij}\,)^2  \,\,.
\label{eq:FREEPHI}
\end{equation}

At the initial time ($t=0$),
the transformed region (product) is simply obtained as a geometrical
deformation of the parent, Fig.\ 1(a), which fixes the initial value of $\phi$ (Fig.\ 1(b)).
Having created this discontinuity $\phi$ at the parent-product interface,
it will diffuse over a time $\tau_{\phi}$. The Langevin equation 
describing the linearised dyamics of the slow modes ${u_i}$ and $\phi$ can be written as,
\begin{equation}
\frac {\rho}{2} {{\ddot u}_i} - \frac {\delta {\cal {F}}} {\delta u_i} =  \phi \nabla_i \frac {\delta {\cal {F}}} {\delta \phi} - \nu_{ijkl} \nabla_j {\dot \epsilon}_{kl}
\label{eq:LANGU}
\end{equation}
\begin{equation}
{\dot \phi}+ {\bf v}\cdot{\bf \nabla}\phi = \frac {1} 
{\tau_{\phi}} {\nabla^2} \frac {\delta {\cal {F}}} {\delta {\phi}} \,\,,
\label{eq:LANGPHI}
\end{equation}
where ${\dot \epsilon}_{ij}$ is the time derivative of the strain tensor $\epsilon_{ij}$.
The inertial term ${\ddot u}_i$ (propagation of sound waves) and solid
viscosity tensor $\nu_{ijkl}$ (damping of sound waves\cite{LL}) are obtained from the coupling of ${\bf u}$ with the momentum density
${\bf g}$ \cite{CHAI}. 
 
We study the consequences of this dynamical theory, in the simple setting
 of a first-order structural transformation from a square (austenite)
 to a rhombic (ferrite) crystal in two dimensions.
Our results, which can be easily extended to the tetragonal
to orthorhombic transition (essentially a 2-dim square to rectangular\cite{KRUM}), are of relevance to structural transformations in alloys like 
In-Pb, In-Tl, Mn-Fe. This transformation involves a shear+volume deformation,
and so the strain order parameter $\epsilon_{ij}$ has only one nontrivial
component $e_3=(u_{xy}+u_{yx})/2$.

We construct a bulk elastic free energy with three minima --- one corresponding to the undeformed square
cell ($e_{3}=0$) and the other two corresponding to the two variants of the rhombic cell ($e_{3}=\pm e_{0}$). The free-energy functional,
in dimensionless variables, is
\begin{equation}
{\cal {F}} = \int_{x, y} a\,{e_3}^2-{e_3}^4+{e_3}^6 
+(\nabla e_3)^2 
  + {\gamma} \,(\phi\, \partial_n e_3)^2
\label{eq:FREE2D}
\end{equation}
The three minima of the homogeneous part of $F$ at $e_3=0$ (austenite)
and $e_3 \equiv \pm e_{0} = \pm [(1+\sqrt{1-3a})/3]^{1/2}$ (ferrite),
 are obtained in the 
parameter range $0 < a < 1/3$. The parameter $a$ is the degree of undercooling $T-T_c$. The surface compressibility $\gamma$ for
an interface between the square ($e_3 = 0$) and the rhombus ($e_3 = e_0$)
positioned along the $y$-axis is given by
$\gamma = (c''(l)+c''(l+e_0l))+c''(|l-e_0l|))/2$, where 
$c''(r)$ is the second
deritive of a typical direct correlation function 
for a two-dimensional fluid whose 
range is taken to be of the order of the distance between next-nearest
neighbours of the parent square lattice of spacing $l$.

There are two time scales relevant to our kinetics --- the quench rate 
${\tau}^{-1} = (dT/dt)/T$ and the 
vacancy relaxation time $\tau_{\phi}$. Accordingly two extreme dynamical
limits suggest themselves.
When $\tau \gg \tau_{\phi}$, the $\phi$ fields relax instantaneously (fast mode) to $\phi = 0$, its equilibrium value. The only surviving slow modes are $u_i$, which
obey Eq.\ (\ref{eq:LANGU}), with the free-energy functional ${\cal {F}} = F_{el}$.
Since we are interested in the growth of the product nucleus in the parent matrix, it is appropriate to rewrite Eq.\ (\ref{eq:LANGU}) in terms of
the broken symmetry mode $e_{3}({\bf r},t)$, which takes the form
$e_{3}({\bf r}-{\bf R}(t))$, when the interface is sharp (${\bf R}(t)$ is
the position of the interface). The equation for $e_3$ is purely
dissipative, ${\dot e}_3 = - \delta F_{el}/\delta e_3$. Minimising $F_{el}$
with respect to $e_3$, subject to boundary conditions $e_3=0$ at $|{\bf r}| \to \infty$ and $e_3 = \pm e_0$ at $|{\bf r}| = 0$, obtains 
a ferrite nucleus, growing as $R \sim t^{1/2}$ at late times\cite{CHAI}.

On the other hand, when $\tau \ll \tau_{\phi}$, the $\phi$ fields are
frozen  in the frame of reference of the nucleating front. In this limit,
$\phi$ is nonzero only at the parent-product interface, and so ${\bf v}$ in
Eq.\ (\ref{eq:LANGPHI}) can be interpreted as the local front velocity.
In the frame front, $\phi$ obeys a diffusion equation with the diffusion coefficient $D_{\phi} = \sqrt {\gamma/\tau_{\phi}}$. Recasting Eq.\ (\ref{eq:LANGU}), in terms of $e_3$, leads again to a purely dissipative equation with
the free-energy functional ${\cal F}$ given by Eq.\ (\ref{eq:FREE2D}).
To determine the structure of the product nucleus, we use a variational
ansastz for $e_3$, consistent with boundary conditions
mentioned above. Consider a rectangular nucleus of length $L$
(along ${\hat x}$) and width $W$ (along ${\hat y}$) 
divided into $N$ twins (Fig.\ 2\,(inset)). The $N-1$ twin interfaces, all of thickness $\eta$, can be parametrized by $e_{3}\,(x)$, which for the $i$-th interface takes values $-e_0$  at  $(i-1)L/N +\eta/2< x < iL/N-\eta/2$ 
and $e_0$ at $iL/N+\eta/2 < x < (i+1)L/N-\eta/2$, connected by
a linear interpolation. The strain at the austenite-ferrite interface
varies linearly between $0$ at  $x < -\xi/2$ and $\pm e_0$ at $x > \xi/2$.
The free energy $E(L,W)$ for the nucleus reads,
\begin{eqnarray}
E(L,W) & =  & [\,\, \Delta F LW + (2I-\Delta F)(L \xi+W \zeta)+ \nonumber \\
        &    & (N-1)(I-\Delta F) W \eta 
+ 2 {e_0}^2 (L/\xi+{W/\zeta})\, +  \nonumber \\
        &    & 4{e_0}^2 (N-1)
{W/\eta}\,\,] + (\gamma {e_0}^4/\xi) [\,\,2 L^3N^{-2}/3 +\nonumber \\
        &    &
L^2 \eta (N^{-2} - N^{-1}) + L {\eta}^2 (1/2 - 3N^{-1}/4)\,\,\, - \nonumber \\
        &    &  L\zeta N^{-1} (LN^{-1} -  \zeta/4 - \eta/2)\,\,] 
+ \gamma {e_0}^4 \zeta^{-1}  \nonumber \\
        &    &  
[\,\, 2W^3/3 - W^2 \xi + W {\xi}^2/2\,\,]
\,\,.
\label{eq:ENERGY}
\end{eqnarray}
In Eq.\ (\ref{eq:ENERGY}), $\Delta F \equiv a {e_0}^2-{e_0}^4+{e_0}^6$ is the difference between the {\it bulk} free energies of the austenite and the ferrite, while $I \equiv a{e_0}^2/3-{e_0}^4/5+{e_0}^6/7$. The structure of the
growing nucleus is obtained by minimising $E(L,W)$
with respect to the order parameter profile $e_3$ (both the amplitude and phase). Within our variational scheme, this amounts to minimising  $E(L,W)$
with respect to the interfacial widths $\eta$, $\xi$ and $\zeta$ and the
phase of the order parameter, $N$. Minimisation yields the structural relation
(Fig.\ 2),
\begin{equation}
\frac {L}{N} \sim W^{\sigma}\,\,.
\label{eq:structure}
\end{equation}
The exponent, $\sigma \sim 1/2$, with tiny deviations for large $W$. 
 The exponent $\sigma$ is empirically known
\cite{KRUM} to lie between $0.4$ and $0.5$. Note that our theory suggests
that Eq.\ (\ref{eq:structure}) {\it holds at all times} during the growth of a martensite 
as a consequence of {\it local equilibrium}. Our prediction for $L/N$ can be
verified from in situ TEM studies of growing martensite
fronts.

Twinning is a consequence of the $\phi$
term which is confined to the parent-product interface. When
 $\tau \ll \tau_{\phi}$, the quench
 nucleates an inclusion which initially grows as a ferrite. Further growth as a single-domain ferrite is discouraged since the energy cost at the interface is
proportional to the square of the discontinuity in ${\bf u}$. The growing nucleus, gets around this by creating a twin. Though this costs interfacial energy due to $(\nabla e_3)^2$ (which is small), the contribution 
from $\phi$ is identically zero at the twin interface, due to symmetry. 

From the contour plots of the optimised free energy plotted in 
Figs.\ 3(a) and 3(b),  
one observes a minimum in $W$ for large
$\gamma \sim {\cal O}(1)$ (Fig.\ 3(b)).
Since we find that, $W_{min} < L$, the free energy
is minimised by thin rectangular strips
reminiscent of `hard' acicular martensites seen in Fe-Ni or Fe-C systems.
 For small $\gamma \sim {\cal O}(10^{-3})$,
there is no such minimum in $W$. As we shall see below, growth along $L$
is much faster than along $W$, and so the martensite traverses the entire extent of the sample along $L$, whereupon growth proceeds along $W$.
Such single interface growth is indeed seen in `soft' solids, like
In-Tl and Au-Cd alloys\cite{SOLID}.  
\myfigure{\epsfysize2.5in\epsfbox{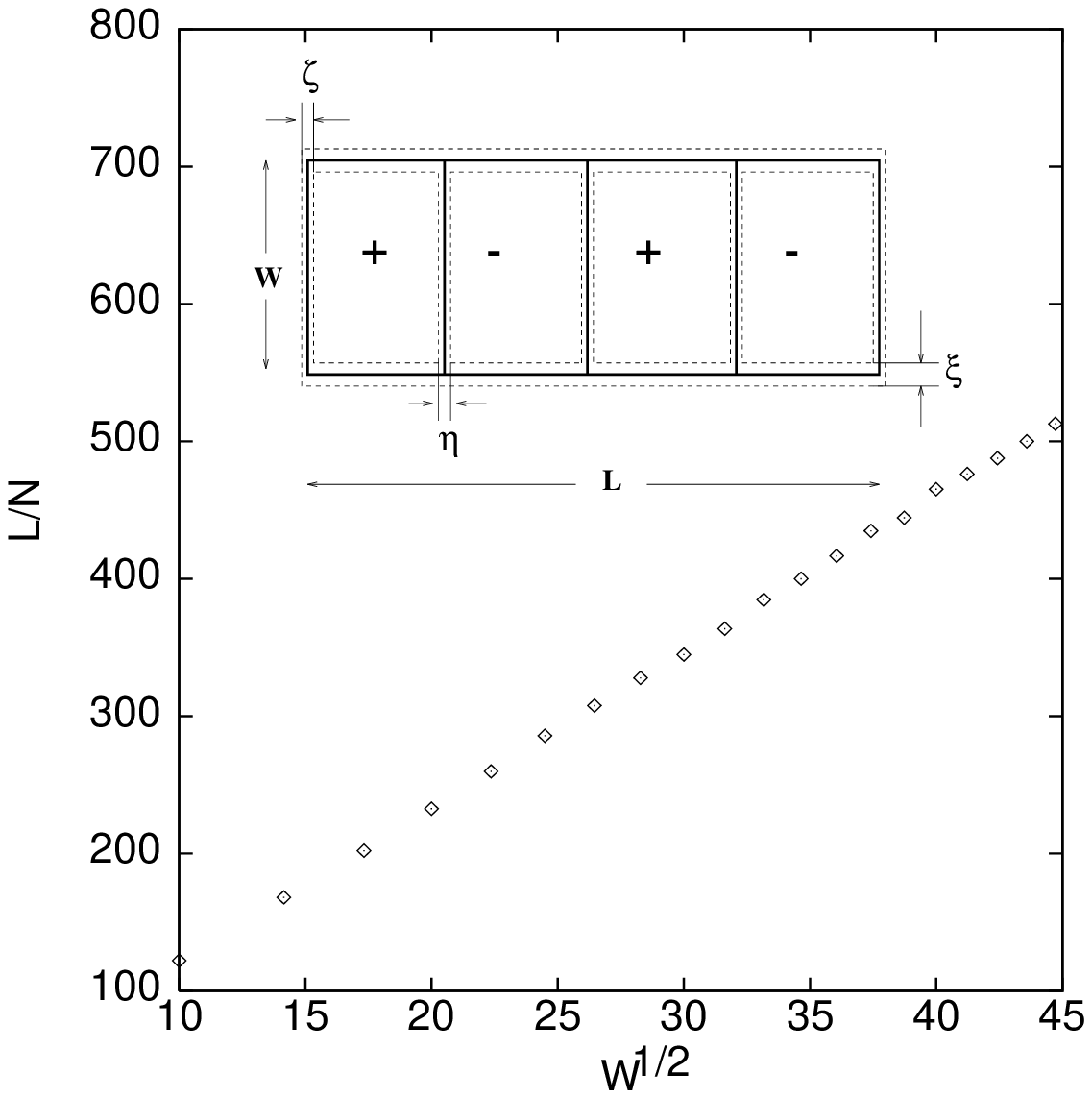}}{\vskip0inFig. \ 2~~
Plot of $L/N$ vs $W^{1/2}$ for a twinned rectangular strip of
length $L$ and width $W$ having $N$ twins, obtained by minimising $E(L,W)$,
for $a=.01$, $\gamma=.001$.
The inset shows the variational profile for $e_3$, with interfacial widths
$\xi$ , $\zeta$ and $\eta$. The $+$ and $-$ denote regions where $e_3=\pm e_0$
respectively.}

The critical nucleus is obtained by locating the saddle point in each of these 
free energy surfaces. Taking the curvature at the corners 
$\kappa = \sqrt {\xi \zeta/LW}$ as a measure of the
sharpness of the growing nucleus, we find the critical nucleus to be
diffuse for both $\gamma = 1$ ($\kappa = 0.4$) and $\gamma = 10^{-3}$ ($\kappa = 1.43$). In situ TEM measurements on critical martensite nuclei, reveal such
diffuse structures\cite{KAMONI}. Once past the critical size, the martensite droplet grows. As growth proceeds
the interfacial widths shrink and $\kappa$ rapidly reduces to a microscopic
value. A well defined front velocity now emerges. We can determine the late
time growth behaviour of the martensite droplet, by requiring that the 
rate of energy change, computed from the time derivative of 
$E(L, W, N \sim L/W^{\sigma})$, equals the energy dissipation from the local
evolution of the order parameter $e_3$ \cite{BRAY}. Since $e_3$ follows a purely dissipative dynamics,
\begin{equation}
 {\dot E}\,(L, W, N \sim L/W^{\sigma}) = \,\nu \int {{\dot e}_3}^2 \,\,dx\, dy\,\,.
\label{eq:DISS}
\end{equation}
The time derivative of $e_3$, can be easily computed from our ansatz, and leads
to the following form for the energy dissipation, $ \nu {e_0}^2 ( {\dot L}^2W/\zeta + {\dot W}^2 L/\xi )$. It is easy to determine the asymptotic
solutions for $L$ and $W$, from the resulting first-order, nonlinear ODE.
Assuming $L(t) = v_L t^{\alpha}$ and  $W(t) = v_W t^{\beta}$ as $t \to \infty$
($\alpha , \beta > 0$), and equating the dominant singular contributions on
either side of Eq.\ (\ref{eq:DISS}), we obtain $\alpha = 1$ and $\beta = 1/2$.
The nonlinear ODE also gives the velocity $v_L \sim \sqrt {\gamma/\nu}$. Since
$\gamma$ is of the order of typical
elastic moduli, the martensite grows with a constant
velocity close to the velocity of sound, in the direction perpendicular
to the twinning plane. Parallel to the twinning planes, the front moves 
diffusively\cite{ONSHON}.
\vspace{-1cm}
\myfigure{\epsfysize3.2in\epsfbox{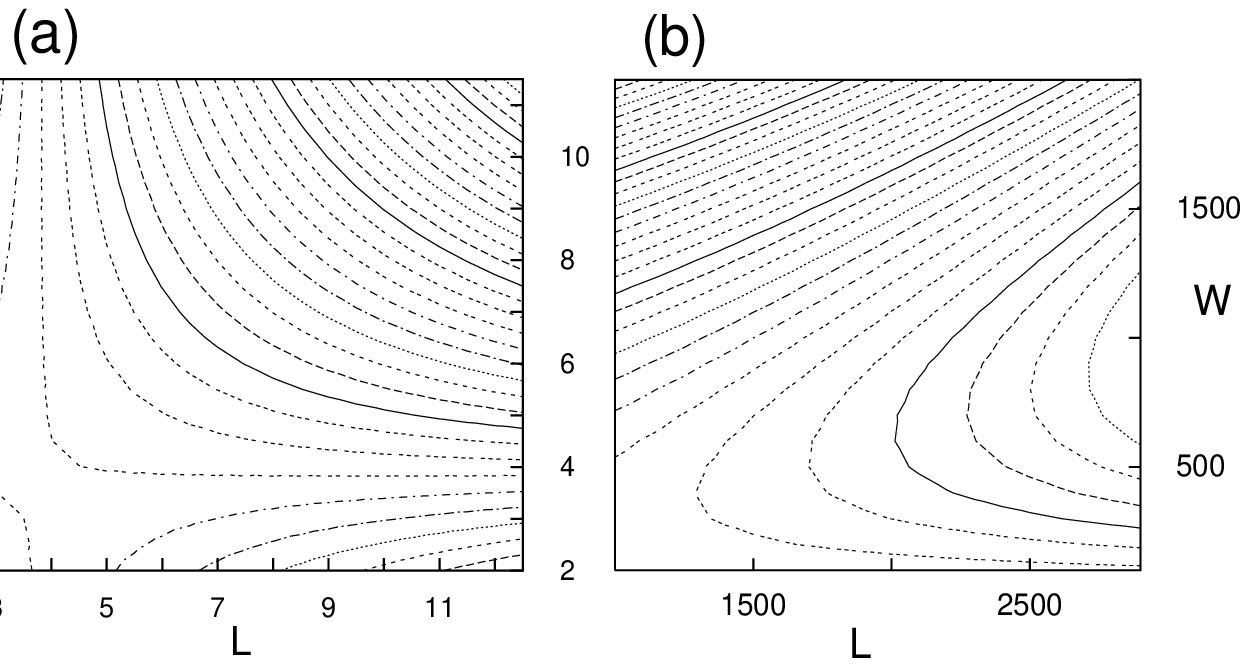}}{\vskip-1inFig.\ 3~~Contour plots of the minimized $E(L,W)$ in the $L-W$ plane for $a=.01$ and 
(a) $\gamma=0.001$ and (b) $\gamma=1$. A comparison of our value for $\eta$
with typical experimental values, yields a unit of length of 
$1 - 2$ {\AA}.}

Thus the martensite nucleus grows in $L$ with velocity $v_L$, till it collides with other growing nuclei, whereupon the high elastic energy barriers prevent coalescence\cite{US}. Growth in $W$ proceeds slowly, upto a point where $W=W_{min}$ (for large $\gamma$). When $\gamma$ is small, growth in $W$ proceeds unimpeded, unless pinned by other nuclei or impurities.
The solid has now got trapped as
a metastable martensite and requires large thermal activation
to transform to the equilibrium ferrite. The martensite can thereafter reduce interfacial energy by deforming the crystal such that the discontinuity in
$\bf u$ (and hence $\phi$) vanishes. This will, however, generate long-range stress fields\cite{KRUM}. In addition, growth can terminate, if $\phi$ is made to relax (say, by raising $T$), leading to thermal arrest of martensites \cite{NISHROIT}.

A more detailed account will explore a `morphology phase-diagram' as a
function of kinetic and structural parameters. In this paper, we have confined ourselves to a simple variational ansatz for the shape.
 Indeed, hard martensites like Fe-Ni are
`lens shaped' plates. We see however, that the interfacial energy at either end of our rectangular strip can be further reduced by decreasing 
$W(x=L)$ and $W(x=0)$, thus forming a `lens'. In three dimensions, the 
possibility of several twin variants would lead to interesting stacking patterns. Moreover, the fascinating
phenomenon of shape memory\cite{NISHROIT} can be studied by
coupling $\epsilon_{ij}$ to an external stress field.

Conventional analysis of the kinetics of martensites\cite{KRUM,GOOD}, 
minimises the elastic
energy, subject to the condition that ${\bf u}$ is continuous across the
parent-product interface. This ad-hoc boundary constraint gives rise to long-range stress fields. The analysis of the dynamics is complicated by the
fact that the boundary constraints are moving, allowing for analytic solutions
only in 1-dim\cite{GOOD}. Our approach does not suffer from this handicap,
since the $\phi$ field, which lives at the interface, naturally leads
to the boundary constraint at late times.

Before we close, we would like to point out some important features not
included in our theory. When the surface compressibility $\gamma$ is large, the product might prefer to generate dislocations at the parent-product interface, 
producing instead internal slip bands\cite{NISHROIT}. Moreover
for solids with low thermal conductivity, transport of heat across the interface might significantly alter the shape of grains. We are currently working on these refinements to the theory.

We thank S.\ G.\ Mishra and T.\ V.\ Ramakrishnan for discussions and encouragement.

\end{document}